\documentclass[letterpaper]{jpconf}
\usepackage{graphicx}

\newcommand{\be}{\begin{equation}}
\newcommand{\ee}{\end{equation}}
\newcommand{\beq}{\begin{eqnarray}}
\newcommand{\eeq}{\end{eqnarray}}

\def\nue{\mathrel{{\nu_e}}}
\def\numu{\mathrel{{\nu_\mu}}}
\def\nutau{\mathrel{{\nu_\tau}}}

\def\barnue{\mathrel{{\bar \nu}_e}}
\def\barnumu{\mathrel{{\bar \nu}_\mu}}
\def\barnutau{\mathrel{{\bar \nu}_\tau}}

\def\t13{\mathrel{{\theta_{13}}}}
\def\y12{\mathrel{{\tan^2 \theta_{12}}}}
\def\c2{\mathrel{{\chi^2 }}}

\newcommand{\df}{DSN$\nu$F}
\newcommand{\snr}{SNR}
\newcommand{\sfr}{SFR}

\begin{document}
\title{The diffuse supernova neutrino flux}

\author{Cecilia Lunardini}

\address{Institute for Nuclear Theory and University of Washington, Seattle, WA 98195 }

\ead{lunardi@phys.washington.edu}

\begin{abstract}
I review the status and perspectives of the research on the diffuse
flux of (core collapse) supernova neutrinos (\df).  In absence of a
positive signal, several upper bounds exist in different detection
channels. Of these, the strongest is the limit from SuperKamiokande
(SK) of 1.2 electron antineutrinos$~{\rm cm^{-2} s^{-1}}$ at 90\%
confidence level above 19.3 MeV of neutrino energy.
%Other upper limits in different channels for electron
%neutrinos and antineutrinos are in the range of $\sim$5-200 ${\rm
%cm^{-2} s^{-1}}$, while the limits on non-electron species are much
%looser.  
The predictions of the \df\ depend on the cosmological rate of
supernovae and on the neutrino emission in a individual supernova
(spectrum, luminosity,..).  Above the SK threshold, they range between
0.05 electron antineutrinos$~{\rm cm^{-2} s^{-1}}$ up to touching the SK limit.
The SK upper bound constrains part of the parameter space of the
supernova rate -- and indirectly of the star formation rate -- only in
models with relatively hard neutrino spectra,
%(average energy of 6.5-8 MeV), 
while predictions with softer spectra would need bounds stronger by
about a factor of $\sim$4 to be tested.  Experimentally, a
feasible and very important goal for the future is the improvement of
background discrimination and the resulting lowering of the detection
threshold.  Theory instead will benefit from reducing the
uncertainties on the supernova neutrino emission (either with more precise
numerical modeling or with data from a galactic supernova) and on the
supernova rate. The latter will be provided precisely by next
generation supernova surveys up to a normalization factor.  Therefore, the detection of the \df\ is likely to be precious
chiefly to constrain such normalization and to study the physics of
neutrino emission in supernovae.
\end{abstract}

\section{Introduction}
\label{intro}
What are our chances to detect neutrinos from core collapse supernovae in the near future?  With current and upcoming neutrino telescopes, a high statistics signal is possible if a supernova occurs in our immediate galactic neighborhood. Such event would be as exciting as it is rare: indeed it could require decades of waiting time, since the rate of core collapse in our galaxy is as low as 1-3 supernovae per century (see e.g. \cite{Arnaud:2003zr,Ando:2005ka}.
A different option is to look for the flux of neutrinos from all supernovae, i.e., integrated over the whole sky. Recently it was shown that the detection of this diffuse supernova neutrino flux (\df) is a concrete possibility \cite{Malek:2002ns}, which, if realized, could turn the field of supernova neutrinos from the realm of rare events to the territory of a moderately paced and steady progress.

Aside from practical advantages, the study of the \df\ has an interest of
its own, because it would give complementary information, on
supernovae and on neutrinos, with respect to an individual supernova
burst.  Since it contains contributions from several supernovae at
different distance and of different morphology, the \df\ reflects the
supernova population of the universe. Thus, from it we could learn
about the distribution of core collapse supernovae with the redshift
and with the mass of the progenitor.  It is known that the supernova
rate (SNR) increases with the redshift: supernovae were more numerous
in the past than at present, so that as much as $\sim 40\% $ of the
\df\ above the SuperKamiokande detection threshold of 19.3 MeV come from cosmological
sources, with redshift $z>0.5$. 
%\cite{}.  
The distribution in mass
goes roughly as the power -2.3 of the progenitor mass, meaning that
about 60\% of the \df\ is produced by relatively small stars with mass
between lower cutoff of 8 $M_\odot$ (the minimum mass to have core
collapse) and $15 M_\odot$, with $M_\odot=1.99\cdot 10^{30}$ kg being the mass of the
Sun. Thus, data from the diffuse flux would complement those we
already have from SN1987A, which had a $\sim15-20 M_\odot$ progenitor.

By testing the SNR, the \df\ also probes, indirectly, the history of
star formation.  Indeed, the SNR is proportional to the star formation
rate (SFR), because supernovae progenitors have a very short lifetime, only
$\sim 10^7$ years %(\ceci{ check!}) 
(three orders of magnitude shorter
than the Sun's lifetime), negligible with respect to their formation
time. Specifically, neutrinos would be  precious to learn
about the normalization of the SNR, since they are not affected by
dust extinction, in contrast with electromagnetic probes.  The diffuse
flux also offers the theoretical possibility to study the first
(Population III) stars \cite{iocco}, since these are believed to have died as core
collapse supernovae, and therefore to have contributed to the \df
\footnote{ The contribution of the first stars would be very hard to
detect, since it accumulates in the lowest energy part of the
spectrum, where the background dominates. }.  

Similarly to a neutrino burst from an individual galactic source, a
detected signal from the \df\ would provide a large amount of
information on the physics of neutrino production, propagation and
emission from a supernova. In particular, such signal would add to the
SN1987A data in constraining numerical models of supernovae, in testing
neutrino oscillations and in probing various effects of physics beyond
the standard model such as the existence of new particles and/or new
forces.  In this paper I will highlight the aspects that are
distinctive the diffuse flux, and refer to other contributions in
these proceedings for general discussions of the physics potential of
supernova neutrinos 
\cite{raffeltproc}.

%%%%%%%%%%%%%%%%%%%%%%%%%%%%%%%%%%%%%%%%%%%%%%%%%%%%%%
\section{Experimental status: upper limits}
\label{limits}

\begin{table}[h]
\caption{\label{ex} Summary of the most stringent bounds on the \df\ from currently active detectors, with their confidence level (C.L.).  The limit on the $\nue$ component labeled as ``indirect" proceeds from the SK $\barnue$ limit with considerations of similarity of the detected $\nue$ and $\barnue$ fluxes due to neutrino oscillations in the star \cite{Lunardini:2006sn}.  The result in the channel $\nue - ^{16}{\rm O}$ is given as an interval of limits, corresponding to the range of  neutrino spectra used in the analysis.  The SNO result is also spectrum-dependent: the quoted bound is the median of several 90\% C.L.  limits found with different neutrino spectra.}
\begin{center}
\begin{tabular}{llll}
\br
Experiment,species  & channel & energy interval & upper limit (${\rm cm^{-2} s^{-1}}$)\\
\mr
KamLAND, $\barnue$ \cite{Eguchi:2003gg} & $\barnue + p \rightarrow n + e^+$ & 8.3  $<$ E/MeV $<$ 14.8 &  $3.7 \times 10^2$ (90\% C.L.) \\
SK, $\barnue$  \cite{Malek:2002ns} & $\barnue + p \rightarrow n + e^+$  & $E/$MeV$>$19.3 &  1.2 (90\% C.L.)  \\
SK/indirect, $\nue$  \cite{Lunardini:2006sn} & & $E/$MeV$>$19.3 &  5.5 ($\sim 98\%$ C.L.) \\
SK, $\nue$  \cite{operes} & $\nue + ^{16}{\rm O} \rightarrow ^{16}{\rm F} + e^-$ & $E/$MeV$>$33 &  61-220  ($90\%$ C.L.) \\
SNO, $\nue$  \cite{Aharmim:2006wq} & $\nue + ^2 _1 {\rm H} \rightarrow p + p + e^-$  &22.9  $<$ E/MeV $<$ 36.9   & 70   \\
LSD, $\numu+\nutau$  \cite{Aglietta:1992yk} & $\nu_{\mu,\tau} + ^{12} {\rm C} \rightarrow ^{12} {\rm C}+ \nu_{\mu,\tau}$  &20  $<$ E/MeV $<$ 100   & $3 \cdot 10^7 $ ($90\%$ C.L.) \\
LSD, $\barnumu+\barnutau$  \cite{Aglietta:1992yk} & $\bar \nu_{\mu,\tau} + ^{12} {\rm C} \rightarrow ^{12} {\rm C}+\bar  \nu_{\mu,\tau}$  &20  $<$ E/MeV $<$ 100   & $3.3 \cdot 10^7$  ($90\%$ C.L.) \\
\br
\end{tabular}
\end{center}
\end{table}
\begin{figure}[h]
\includegraphics[width=35pc]{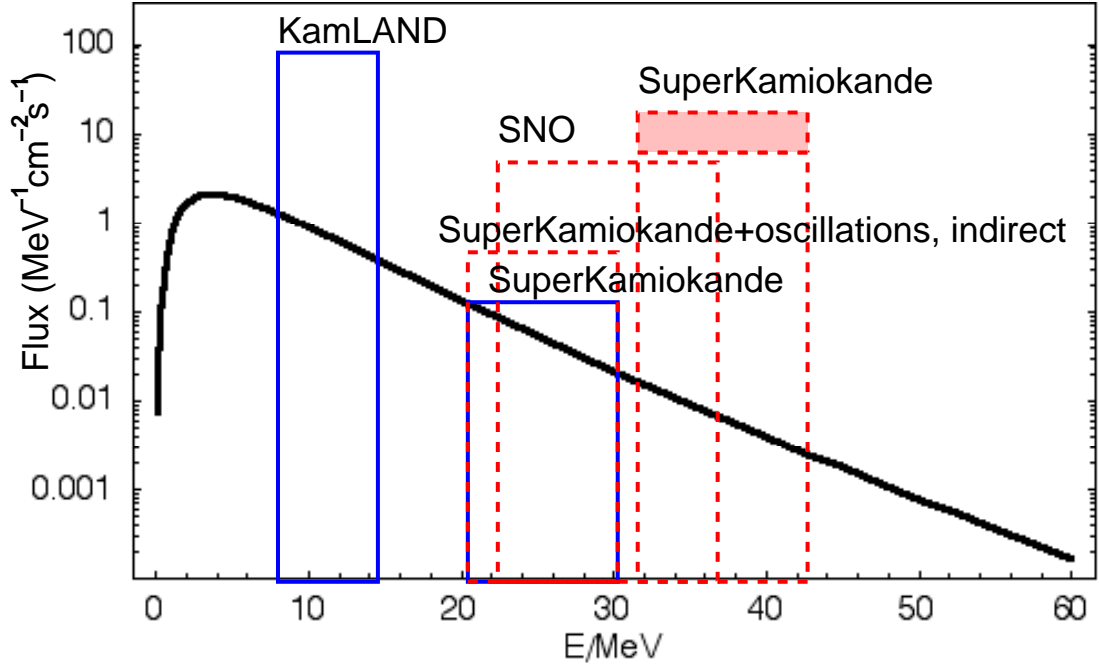}\hspace{2pc}%
%\begin{minipage}[b]{18pc}
\caption{\label{experimental_limits} Graphical view of the bounds in Table \ref{ex}.  The solid (blue) lines and the dashed (red) lines refer to $\barnue$ and to $\nue$ respectively.  The limits are compared with an indicative neutrino spectrum (from ref. \cite{Ando:2004hc}).  To make the comparison possible, each upper bound from the Table \ref{ex} has been divided by the size of the interval of neutrino energy where the search was done. In case of an open interval (as it is the case for SK), a 10 MeV interval has been assumed. A bin of this size is expected to capture practically the whole flux above the threshold, due to the steep (exponential) falling of the neutrino spectrum with energy.   The LSD limit on neutrinos of non-electron flavor (not shown) lies outside the scale of this plot.}
%\end{minipage}
\end{figure}
So far, the \df\ has escaped detection.  In Table
\ref{ex} and fig. \ref{experimental_limits} I summarize the most  stringent bounds  available on this flux.   Thanks to their larger volumes,  currently active detectors detectors \cite{Malek:2002ns,Eguchi:2003gg,Aharmim:2004uf,Aharmim:2006wq} have improved dramatically on the limits set by the previous generation of experiments \cite{Zhang:1988tv,Aglietta:1992yk}.  In particular, the 50 kt of water of SuperKamiokande (SK) has allowed to push the limit on
both the $\barnue$ and $\nue$ components of the \df\
within an order of magnitude or so from theoretical predictions (see
Sec. \ref{theory} for those). 
The SK data have been analyzed by the SK collaboration in the dominant detection channel, inverse beta decay induced by electron antineutrinos \cite{Malek:2002ns}.   The same data have been used by others to constrain the $\nue$ flux by looking for charged-current $\nue$ interactions on $^{16}{\rm O}$ \cite{operes}.  Due to the smaller cross section of this process, the resulting limit on the $\nue$ flux is looser than the result from inverse beta decay, but it is still interesting as it improves substantially on the older result by LSD \cite{Aglietta:1992yk}.  With its 1 kt tank of heavy water, SNO could look for $\nue$ charged current interactions on deuterium, putting an upper limit comparable with the bound from $\nue - ^{16}{\rm O}$ in SK.  It has to be noticed, however, that the best constraint on $\nue$ from the diffuse flux comes from the constraint on the $\barnue$ component at SK, by considering that the two components must be similar due to their common origin  in the non-electron neutrino flavors inside the star through neutrino oscillations \cite{Lunardini:2006sn}.

The data collected by currently active detectors have not yet been analyzed to constrain the non-electron components of the \df. On these, the loose limits put by LSD \cite{Aglietta:1992yk} (see Table \ref{ex}), are still the best available. However,  considerations of naturalness lead to believe that constraints at the level of those on $\nue$ and $\barnue$ should apply to non-electron neutrinos as well.

The main challenge and limiting factor of experimental searches of the diffuse flux is 
background reduction. At a water Cerenkov detector like SK a search for the diffuse flux requires to cut all events with energy below 18 MeV (positron energy), due to the high spallation background below that threshold. This excludes the bulk of
the flux, which is concentrated at lower energy, $\sim 5$ MeV, causing
a huge loss of sensitivity with respect to the ideal case of no
background. 
 In the remaining energy window one has to look for the signal induced by the \df\ ($\barnue$ component) on top of the ineliminable background from invisible muons and atmospheric neutrinos, which limit the sensitivity further.  
 Similar considerations hold for  heavy water.  Liquid scintillator allows to single out inverse beta decay events by observing the positron and neutron capture signals in coincidence. This results in a better background reduction and thus explains the sensitivity of KamLAND down to energy of about 8 MeV, which is where  the ineliminable background of reactor neutrinos ends \footnote{This consideration on the reactor neutrino background motivates the initiative of building a large liquid scintillator detector in a de-nuclearized area such as Hawaii, New Zealand or Australia, see e.g. \cite{wurm}.}.

%\ceci{part written slowly and with boredom... better rewrite?  Better to change structure and go experiment by experiment, instead than overview-channels-background discussion?}

%%%%%%%%%%%%%%%%%%%%%%%%%%%%%%%%%%%%%%%%%%%%%%%%%%%%%%
\section{Status of theory: flux predictions}
\label{theory}

The recipe to estimate the \df\ is relatively simple: consider the
neutrino output of an individual supernova, 
%\footnote{\ceci{write something about individual variations}}
 apply the relevant
propagation effects -- such as redshift of energy and neutrino
oscillations -- and then sum over the supernova population of the
universe.  Formally, this corresponds to the following integral: 
\be
\Phi(E)=\frac{c}{H_0}\int_0^{z_{ max}} R_{ SN}(z)
\sum_{w=e,\mu,\tau} \frac{{d} N_w(E')}{{d} E'} P_{{  w}{ e}}(E, z)
 \frac{{d}
z}{\sqrt{\Omega_{ m}(1+z)^3+\Omega_\Lambda}} ~,
\label{conv}
\ee which describes the $\nue$ component of the flux differential in
the neutrino energy at Earth, $E$.  There $dN_w(E^\prime)/dE^\prime$
is the flux of neutrinos of flavor $w$ emitted by an individual
supernova, differential in the neutrino energy at production,
$E^\prime$. $P_{we}$ is the probability that  a neutrino produced
as $\nu_w$ is detected as $\nue$ at Earth, and $R_{SN}$ describes the
\snr\ per comoving volume. $\Omega_m\simeq 0.3$ and
$\Omega_\Lambda\simeq 0.7$ represent the fractions of energy density
of the universe in matter and dark energy respectively, $c$ is the
speed of light and $H_0\simeq 70~{\rm Km~ s^{-1} Mpc^{-1}}$ is the
Hubble constant.

\begin{figure}[h]
\includegraphics[width=40pc]{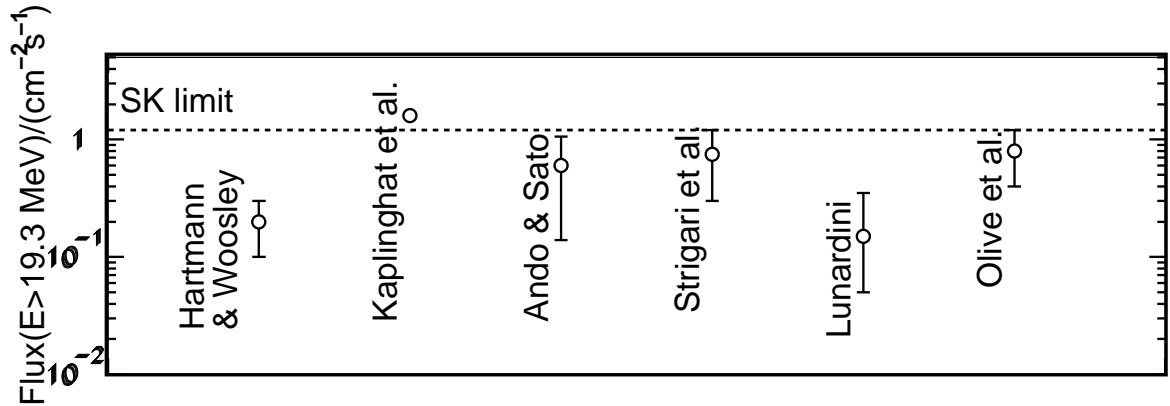}\hspace{2pc}%
%\begin{minipage}[b]{18pc}
\caption{\label{comparefig} A sample of theoretical predictions \cite{Hartmann:1997qe,Kaplinghat:1999xi,Ando:2004sb,Ando:2004hc,Strigari:2003ig,Lunardini:2005jf,Daigne:2005xi} for the $\barnue$ component of the \df\ above the SK energy threshold (figure adapted from ref. \cite{Lunardini:2005jf}).  The number  by Lunardini is quoted with a 99\% C.L. error bar. For the other results, the error bars are an indicative description of the uncertainty due to uncertain input parameters; they have no statistical meaning.  The SK limit (see Table \ref{ex}) is shown for comparison.}
%\end{minipage}
\end{figure}

Many estimates of the \df\ according to Eq. (\ref{conv}) have been
published in the literature \cite{Bisnovatyi-Kogan:1982rd,Krauss:1983zn,woosleyetal,Totani:1995rg,Totani:1995dw,Malaney:1996ar,Hartmann:1997qe,Kaplinghat:1999xi,Ando:2002ky,Strigari:2003ig,Ando:2004sb,Ando:2004hc,Iocco:2004wd,Lunardini:2005jf,Daigne:2005xi}. Fig. \ref{comparefig} shows a sample of
results for the $\barnue$ component of the flux above the SK
threshold, compared with the current SK limit (see Table
\ref{ex}.)   As it appears in the
figure, there is a considerable spread in the flux predictions, which
range -- considering the errors quoted -- roughly between $0.05~{\rm
cm^{-2} s^{-1}}$ and values that even exceed the experimental limit
of $1.2 ~{\rm cm^{-2} s^{-1}}$, thus resulting in constraints on the
input quantities of the calculation (see Sec. \ref{learnt}).  This
spread reflects the different approaches used by different
authors. Specifically, Hartmann and Woosley \cite{Hartmann:1997qe}, Ando and Sato
\cite{Ando:2004sb,Ando:2004hc}, Strigari et al. \cite{Strigari:2003ig} and Olive et al. \cite{Daigne:2005xi}  have
used the \snr\ as it is inferred from measurements of the \sfr, while
Kaplinghat et al. \cite{Kaplinghat:1999xi} have estimated the \snr\ considering the constraints
on the universal metal enrichment history.  Lunardini
\cite{Lunardini:2005jf} used information on the \snr\ from direct supernova
observations only.  Different were also the choices of the neutrino
spectrum: all the references take the spectra from numerical
simulations, with the exception of ref. \cite{Lunardini:2005jf}, where only the
(softer) spectra that fit the SN1987A data are considered, following
the earlier example of Fukugita and Kawasaki \cite{Fukugita:2002qw}.

Predictions exist also for the flux of $\barnue$ above energy
thresholds lower than the current SK one. These could be relevant for
future neutrino telescopes, e.g. SK with Gadolinium addition (see
Sec. \ref{discu}). Table \ref{pred} shows the results
obtained in ref. \cite{Lunardini:2005jf} for the $\barnue$ flux and the corresponding
rate of events from inverse beta decay in SK. From it, one concludes
that a lowering of the energy threshold down to $\sim 10$ MeV would
represent a large improvement in the flux captured, with no guarantee
of a detection, however, due to the smallness of the number of events. 

%\begin{minipage}[b]{18pc}
\begin{table}
\caption{\label{pred} Predictions for the $\barnue$ diffuse flux and for the corresponding rate of inverse beta decay events in SK for different energy thresholds, from ref. \cite{Lunardini:2005jf}.  The intervals correspond to 99\% C.L..  These results were obtained using  soft neutrino spectra compatible with the SN1987A data, and therefore lie on the conservative side with respect to predictions that used harder neutrino spectra, motivated by numerical simulations. }
\begin{center}
\begin{tabular}{llll}
\br
 & $E>19.3$ MeV & $E>11.3$ MeV & $E>5.3$ MeV \\
\mr
flux (${\rm cm^{-2} s^{-1}}$)  & 0.05 - 0.35   & 0.33 - 2.1  & 3.2 - 22.6 \\
events/year at SK  & 0.09 - 0.7  & 0.27 - 1.6 & 0.43 - 2.2 \\
\br
\end{tabular}
\end{center}
\end{table}
%\end{minipage}

%%%%%%%%%%%%%%%%%%%%%%%%%%%%%%%%%%%%%%%%%%%%%%%%%%%%%%
\section{What have we learned on supernovae and on neutrinos? What will we learn?}
\label{learnt}

\begin{figure}[h]
\includegraphics[width=26pc]{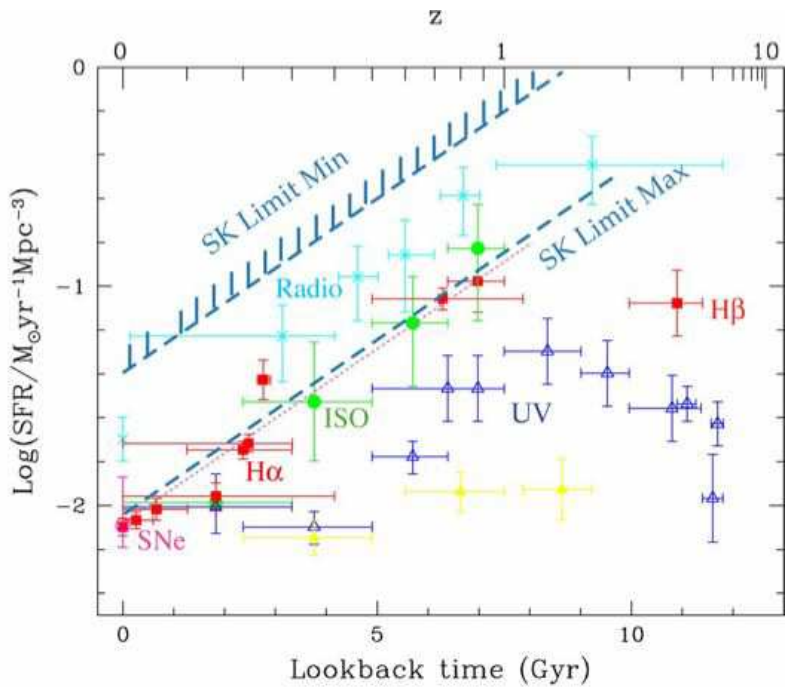}\hspace{2pc}%
%\begin{minipage}[b]{24pc}
\caption{\label{fukugitafig} Measurements of the \sfr\ compared with the exclusion region obtained from the SK limit on the $\barnue$ diffuse flux (see Table \ref{ex}), from ref. \cite{Fukugita:2002qw}.  The figure  shows how the exclusion region changes by varying the input neutrino spectrum in the range allowed by SN1987A.  The less stringent exclusion barely touches the astrophysical measurements.}
%\end{minipage}
\end{figure}

Undoubtedly, the best piece of information that we have, at present,
on the \df\ is the negative result of SK. Is this upper limit strong
enough to give any information? The answer can be read off from
fig. \ref{comparefig}: there one can see that the SK limit touches some of the
theoretical predictions but not others, with the conclusion that only
conditional bounds can be put on the \snr\ (or, indirectly, on the
\sfr) or on the neutrino emission in a supernova.  The situation is
illustrated well in fig. \ref{fukugitafig}, taken from ref. \cite{Fukugita:2002qw}.  The
figure shows how the exclusion region for the \sfr\ varies by varying
the neutrino spectrum in the region allowed by SN1987A and the minimum
progenitor mass between 8 and 10 $M_\odot$.  The space allowed by
astrophysical measurements of the \sfr, also shown in the figure, is
only marginally touched by the most conservative exclusion line,
corresponding to the softest neutrino spectrum and 10 $M_\odot$
minimum progenitor mass (``SK Limit Min" in the figure). Some
restriction of this space is obtained if the hardest spectrum is used
with 8 $M_\odot$ minimum progenitor mass.  The conclusions of
ref. \cite{Lunardini:2005jf} are analogous. The exclusion found by Strigari et
al. \cite{Strigari:2005hu} is not in constrast with what shown in fig. \ref{fukugitafig}, since
these authors relied on a harder neutrino spectrum, with respect to refs. \cite{Fukugita:2002qw} and \cite{Lunardini:2005jf}, motivated
by numerical simulations.

On the side of neutrino physics, the \df\ could be the best probe of
exotic effects that could manifest themselves only on cosmological
distances. An example of this is neutrino decay: it was shown \cite{Ando:2003ie,Fogli:2004gy}
that data from the diffuse flux would be sensitive to a ratio of
neutrino lifetime over mass as large as $\tau/m \simeq 
10^{10}~{\rm s/eV}$.  It was also pointed out \cite{Goldberg:2005yw} that the \df\
could reveal the existence of new, light gauge bosons that could be
produced in the resonant annihilation of a neutrino and an
antineutrino, one of them from a supernova and the other from the
cosmological relic neutrino background.  A test of dark energy,
complementary to astrophysical measurements, is also possible in
principle \cite{Hall:2006br}.  Many other aspects of the physics of neutrinos and supernovae
could be tested with the \df, similarly to the case of an individual
supernova burst. These are illustrated in detail elsewhere in these proceedings \cite{raffeltproc}.

%\ceci{add about testing the dark energy? Add more... can we learn only 2-3 things? Say that here I mention only things that are distinctive of the \df, and refer to other talks for less distinctive aspects.}

%%%%%%%%%%%%%%%%%%%%%%%%%%%%%%%%%%%%%%%%%%%%%%%%%%%%%%
\section{Discussion: perspective of future research}
\label{discu}

What are the likely developments in the study of the \df\ in the next
5-10 years?

Progress will be made with new, more sensitive neutrino
telescopes. The first to become operational could be the GADZOOKS
project \cite{Beacom:2003nk}: an upgraded configuration of SK
employing a solution of water and Gadolinium trichloride instead of
the pure water like the present detector. The presence of Gadolinium
would greatly enhance neutron capture, resulting in better background
discrimination for the search of the $\barnue$ component of the \df.
This would make it feasible to lower the energy threshold to 11.3 MeV
in neutrino energy (10 MeV in positron energy) and therefore to have a
larger event rate, estimated to be as high as 2
\cite{Lunardini:2005jf} or even 6 \cite{Beacom:2003nk} events/year.
The GADZOOKS initiative is progressing with the creation of a
dedicated committee internal to the SK collaboration and the
construction of a new test tank made of stainless steel instead than
coated carbon steel as the previously used K2K 1 kt prototype, where
rust made testing impossible \cite{vaginssbovoda}.
%\ceci{(check svoboda's talk at NNN06)}

In the space of a decade from now, water Cerenkov detectors of
megaton mass, 20 times larger than  SK, could become a
reality.  Projects of this type are under study: these are
HyperKamiokande \cite{Nakamura:2003hk}, UNO \cite{Jung:1999jq} and MEMPHYS \cite{Mosca:2005mi}. According to
a very conservative estimate \cite{Lunardini:2005jf}, these should have an event rate
between 2 and 44 events/year above the current SK threshold of 19.3
MeV.

A rather intense activity is ongoing to plan large non-water
neutrino detectors. One of these is LENA \cite{MarrodanUndagoitia:2006re}, which, with its 50 kt
of liquid scintillator, would have a better background rejection than
SK and a comparable event rate.   Detectors using $\sim$ 100 kt of liquid Argon, like GLACIER  \cite{glacier} and LANNDD  \cite{Cline:2006st} would be precious for their sensitivity to the $\nue$ component of the \df. 
%\ceci{Add details? Number of events?}

On the side of theory, much work has to be done to improve the
predictions of the \df. 

  To reduce the uncertainty on the estimated diffuse flux, it would be
crucial to reduce the uncertainties on the neutrino fluxes and spectra
emitted by an individual supernova. This could be offered by the
advancement of numerical simulations or by data from a future galactic
supernova.
Besides knowing better the neutrino emission by a single star, it would be important  to generalize the calculation of the diffuse flux,
Eq. (\ref{conv}), to include individual variations of the neutrino output
between different stars, depending on various factors like the
progenitor mass, rotation, magnetic fields, etc..
The uncertainty on the diffuse flux associated with the \snr\  will  be dramatically reduced when results become available from the next generation supernova surveys like SNAP \cite{SNAP} and JWST \cite{jwst}. While primarily designed to study type Ia supernovae, these would see thousands of core collapse supernovae up to redshift $\sim 1$ and beyond \cite{SNAP}.

At the interface between theory and experiment, work is needed to
improve the interpretation of the existing experimental searches: it
is necessary to consistently take into account that a bound on the
neutrino flux from an experimental limit on the event rate necessarily
depends on the neutrino spectrum, which is not known precisely at the
present time.

Finally, let us review the scenarios that could be realized with new
experimental results on the $\barnue$ component of the \df\ and the
current theoretical predictions as in fig. \ref{comparefig}.  Evidence of the
diffuse flux above the SK limit would point in the direction of a
neutrino spectrum much harder than what used in the analysis of the
current SK data \cite{Malek:2002ns}, or would indicate a fluctuation in the flux
due to an extragalactic supernova at moderate distance \cite{Ando:2005ka}. The
latter case could be distinguished on the basis of the time
distribution of the excess flux.  A detection of the neutrino flux
anywhere below the SK limit would be very important to discriminate
between the different predictions. To  constrain the \snr\
unambiguously (i.e., to obtain a constraint  in every
framework of theory considered so far), would require upper limits on the diffuse $\barnue$ at the level of $\sim 0.3~{\rm cm^{-2} s^{-1}}$ above the current SK threshold.

\ack
I acknowledge support in my research and travels from the  INT-SCiDAC grant number DE-FC02-01ER41187. I warmly thank the organizers and the participants of the Neutrino 2006 conference for the efficient organization and the intellectually fertile atmosphere I enjoyed there.

\section*{References}

\bibliography{ceciproc}
\bibliographystyle{apsrev}

%%%%%%%%%%%%%%%%%%%%%%%%%%%%%%%%%%%%%%%%%%%%%%%%%%%%%%%
%\begin{\thebibliography}{99}
%\item Strite S and Morkoc H 1992 {\it J. Vac. Sci. Technol.} B {\bf 10} 1237 
%\item Jain S C, Willander M, Narayan J and van Overstraeten R 2000 
%{\it J. Appl. Phys}. {\bf 87} 965 

%\end{\thebibliography}
%%%%%%%%%%%%%%%%%%%%%%%%%%%%%%%%%%%%%%%%%%%%%%%%%%%%%%
%\section{ }
%\label{ }
%%%%%%%%%%%%%%%%%%%%%%%%%%%%%%%%%%%%%%%%%%%%%%%%%%%%%%
%\section{ }
%\label{ }

\end{document}